\documentclass[11pt]{article}

\usepackage{graphicx,amsmath,amsfonts,amssymb,dcolumn,amsthm,slashed,bbm,xspace,pgffor,tikz,mathbbol,latexsym,enumerate,amsmath}
\usepackage{soul}
\usepackage[colorlinks,citecolor=red,urlcolor=cyan,linkcolor=blue]{hyperref}
\usepackage[utf8]{inputenc}
\usepackage[english]{babel}
\usepackage{lmodern}
\usepackage{microtype}
\usepackage{cite}

\numberwithin{equation}{section}

\begin{document}

\title{\textbf{Unifying fractons, gravitons and photons from a gauge theoretical approach}}

\author{\textbf{Rodrigo F.~Sobreiro}\thanks{rodrigo\_sobreiro@id.uff.br}\\\\
\textit{{\small UFF - Universidade Federal Fluminense, Instituto de F\'isica,}}\\
\textit{{\small Av. Litorânea, s/n, 24210-346, Niter\'oi, RJ, Brasil.}}}

\date{}
\maketitle

\begin{abstract}
We revisit the first principles gauge theoretical construction of relativistic gapless fracton theory developed by A.~Blasi and N.~Maggiore. The difference is that, instead of considering a symmetric tensor field, we consider a vector field with a gauge group index, \emph{i.e.} the usual Einstein-Cartan variable used in the first order formalism of gravity. After discussing the most general quadratic action for this field, we explore the physical sectors contained in the model. Particularly, we show that the model contains not only linear gravity and fractons, but also ordinary Maxwell equations, suggesting an apparent electrically charged phase of, for instance, spin liquids and glassy dynamical systems. Moreover, by a suitable change of field variables, we recover the Blasi-Maggiore gauge model of fractons and linear gravity.
\end{abstract}

\maketitle

\section{Introduction}\label{intro}

In recent years, collective excitations called fractons\footnote{It is worth mentioning that the notion of fractons discussed in this paper is different from the one sometimes appearing in quark confinement context \cite{Khlopov:1981wm,Khlopov:1999rs,Khlopov:2012lua}.} have been drawing considerable attention in solid state physics, see for instance \cite{Vijay:2016phm,Nandkishore:2018sel,Pretko:2020cko}. These types of quasi-particles are supposed to exist in, for example, spin liquids \cite{Nandkishore:2018sel,Pretko:2020cko} and glassy dynamics \cite{Prem:2017qcp,Nandkishore:2018sel,Pretko:2020cko}. Fractons have also been connected to elasticity theory \cite{Pretko:2017kvd,Nandkishore:2018sel,Pretko:2020cko} and gravity \cite{Pretko:2017fbf,Nandkishore:2018sel,Pretko:2020cko}. In a gauge theoretical description \cite{Pretko:2016lgv,Pretko:2020cko,Bertolini:2022ijb,Blasi:2022mbl,Bertolini:2023juh}, fractons carry fractonic charge and produce (and interact through) $U(1)$ tensor fields which generalize the usual Maxwell electromagnetism \cite{Maxwell:1865zz,Jackson:1998nia}. In such fractonic electrodynamics, there are two types \cite{Pretko:2016lgv}: one with scalar fractonic charges where not only charges but also dipole moment and one specific component of the quadrupole moment are conserved; and another with vector charges where charges and its angular momentum analogue are conserved. For instance, for vector charge theory, if $E^{ij}=E^{ji}$ stands for the tensor electric field and $\rho^i$ for vector fractonic charge density, Gauss law reads\footnote{Latin indices from $i$ to $q$ run through $\{1,2,3\}$. We also define Greek indices to run through $\{0,1,2,3\}$ while Latin indices from $a$ to $h$ run through $\{\underline{0},\underline{1},\underline{2},\underline{3}\}$.}
\begin{equation}
    \partial_j E^{ij}=\rho^i\;,\label{frac1}
\end{equation}
whose natural covariant generalization \cite{Bertolini:2022ijb} yields
\begin{equation}
    \partial_\alpha F^{\mu\nu\alpha}=\tau^{\mu\nu}\;,\label{frac2}
\end{equation}
with $F^{\mu\nu\alpha}=F^{\nu\mu\alpha}$ being the fracton field strength (See Section \ref{BBDM2}) and $\tau^{\mu\nu}=\tau^{\nu\mu}$ the generalized fractonic current density. Indeed, it was shown in \cite{Bertolini:2022ijb} that equation \eqref{frac2} accommodates a complete covariant generalization of vector fractonic electrodynamics. 

Just like ordinary electrodynamics, fractonic electrodynamics can be formulated in terms of potentials and gauge symmetry \cite{Pretko:2016lgv,Pretko:2020cko,Blasi:2022mbl,Bertolini:2022ijb,Bertolini:2023juh}. Moreover, a gravitational parallel between gravity and fractons was established in \cite{Pretko:2017fbf}. In fact, as demonstrated by Blasi and Maggiore \cite{Blasi:2022mbl}, the construction of a theory of fractons based on the gauge principle, produces an action composed by the usual fracton action and the Fierz-Pauli action for spin-2 massless gravitons \cite{Fierz:1939ix,Gambuti:2021meo}. Therefore, the covariant Blasi-Maggiore (BM) model generalizes pure fractonic vector gauge theory to include gravitational modes on it. See also \cite{Bertolini:2023juh}. Later, Bertolini and Maggiore found their way to formulate the BM model in terms of the fracton field strength appearing in equation \eqref{frac2}. 

The BM model is constructed with a symmetric tensor field $h_{\mu\nu}$, transforming under gauge transformations of the form
\begin{equation}
    \delta_f h_{\mu\nu}=\partial_\mu\partial_\nu\zeta\;,\label{gt1}
\end{equation}
with $\zeta$ being a local parameter. It turns out that transformation \eqref{gt1} is just a special case of the gauge transformation in linear gravity theories (linear diffeomorphisms) \cite{Wald:1984rg}, namely
\begin{eqnarray}
    \delta_d h_{\mu\nu}=\partial_\mu\zeta_\nu+\partial_\nu\zeta_\mu\;,\label{diff1}
\end{eqnarray}
for the restricted set of longitudinal diffeomorphisms\footnote{It is interesting that, in the search of quantum gravity theories, unimodular gravity is quite popular. In this case, longitudinal diffeomorphisms are disregarded from the beginning in favor of the transverse ones. See, for instance, \cite{Unruh:1988in,Eichhorn:2013xr} and references therein. Remarkably, fractonic physics emerges right from the disposals of unimodular gravity.} characterized by $\zeta_\mu=\partial_\mu\zeta/2$. Therefore, because the Fierz-Pauli is invariant under the longitudinal diffeomorphism in expression \eqref{diff1}, it is obviously also invariant under transformations \eqref{gt1}. The main conclusion of their gauge approach is that gravitonic modes are inherent to fracton models due to the fact that transformations \eqref{gt1} are a special case of the linear diffeomorphism transformations \eqref{diff1}. But the contrary is not true because the fractonic action is not invariant under the full diffeomorphic transformations \eqref{diff1}.

In the present paper, we construct a gauge fractonic model using Einstein-Cartan (EC) linear variables \cite{Utiyama:1956sy,Kibble:1961ba,Sciama:1964wt,Misner:1973prb,DeSabbata:1986sv,Mardones:1990qc,Zanelli:2005sa}. For simplicity, and for the sake of comparison with the BM model \cite{Blasi:2022mbl}, we neglect the independent sector of the spin-connection in this work. The action we gain is indeed equivalent to the one in the BM model, as we prove by a simple change of field variables. One useful feature in working with EC variables in linear gravity is that we can work with two distinct sets of indices, both transforming under Lorentz transformations. But one set of indices refers to spacetime itself while another set refers to the local tangent space. Thence, by fixing a regime to our model where frames in tangent space do not move with respect to each other (all observers have synchronized clocks), the model reveals a sector which is indistinguishable from ordinary electromagnetism. In this sense, the BM gauge theory of fractons embraces not only fractons and graviton-like modes, but also photon-like excitations. And this feature is made evident in terms of the EC variables.

This work is organized as follows: In Section \ref{BBDM} we construct the gauge model of fractons in EC variables. In Section \ref{MAX} we discuss the regime of the model where fracton electrodynamics behave exactly as ordinary Maxwell theory. In Section \ref{BBDM2} it is shown that the model considered in this work coincides with BM model of fractons. Finally, In Section \ref{conc} we display our final considerations.

\section{Gauge theory of a special kind}\label{BBDM}

To begin with the construction of our model, we set the scenario to be the Minkowski spacetime with metric $\eta\equiv\mathrm{diag}(+1,-1,-1,-1)$. We define the Minkowskian vierbein $\eta^a_\mu$ (and its inverse $\eta^\mu_a$) to be the set of maps connecting local inertial frames $x^a$ to world coordinates $x^\mu$, namely $x^a=\eta^a_\mu x^\mu$ and $x^\mu=\eta^\mu_ax^a$. Because the whole spacetime is flat, all frames are therefore taken to be inertial. Geometrodynamically speaking, this means that no gravitational effects are in place, \emph{i.e.}, spacetime is not dynamical. For simplicity, $\eta^a_\mu$ is thus assumed to carry constant components and all coordinate systems are taken as Cartesian ones. In the EC formalism of gravity \cite{Utiyama:1956sy,Kibble:1961ba,Sciama:1964wt,Misner:1973prb,DeSabbata:1986sv,Mardones:1990qc,Zanelli:2005sa}, the vierbein $e^a_\mu$ locally maps general frames $dx^\mu$ to inertial ones $dx^a$. The inertial frames are usually taken as Cartesian, but they could be curvilinear as well. Hence, a more general approach in flat space could consider curvilinear yet inertial frames in both sides of the vierbein map. However, this is an unnecessary complication we avoid. 

The fundamental field of the model is $h^a_\mu$, usually identified with the graviton field in linear gravity theories. In here, this is precisely the general fracton potential field. Together with $h^a_\mu$, the field $h_a^\mu$ is also important for covariance and the following relations are valid:
\begin{eqnarray}
\eta^a_\mu\eta^\mu_b&=&\delta^a_b\;,\nonumber\\
    \eta^a_\mu\eta^\nu_a&=&\delta^\nu_\mu\;,\nonumber\\
    h^\mu_a&=&\eta_b^\mu\eta^\nu_a h^b_\nu\;.\label{etah1}
\end{eqnarray}
Naturally, Latin and Greek indices transform under global Lorentz transformations\footnote{All results attained in the present work are also achieved by considering the Poincaré group as well.}. We also define linear diffeomorphisms, a genuine local Abelian gauge transformation, in the form
\begin{eqnarray}
\delta_dh^a_\mu&=&\frac{1}{2}\eta^a_\nu\left(\partial_\mu\zeta^\nu+\partial^\nu\zeta_\mu\right)\;,\nonumber\\
\delta_dh_a^\mu&=&\frac{1}{2}\eta_a^\nu\left(\partial^\mu\zeta_\nu+\partial_\nu\zeta^\mu\right)\;.\label{diff2}
\end{eqnarray}
In the EC formalism, besides the graviton, the spin-connection is also considered as a fundamental field and independent from $h^a_\mu$. Nevertheless, this field is not important for our current goals and we do not consider it here. 

Let us perform two exercises. The first one is to find the most general quadratic theory describing the dynamics of the graviton. This means to find the most general action, respecting the global Lorentz symmetries and the gauge symmetry \eqref{diff2}, polynomially local on the fields and their derivatives and quadratic on $h^a_\mu$. We also consider that the action depends mostly quadratically on the derivatives. Moreover, $h^a_\mu$ is taken as a massless field from the beginning. The most general possible action with arbitrary coefficients $a_I$, already respecting global Lorentz symmetries, is just
\begin{equation}
    S^h=\int d^4x\left(a_1h\Box h+a_2h^a_\mu\Box h^\mu_a+a_3h\eta^\mu_a\partial_\mu\partial^\nu h^a_\nu+a_4h^\mu_a\partial_\mu\partial^\nu h^a_\nu\right)\;,\label{acth1}
\end{equation}
with $h=\eta_a^\mu h_\mu^a$ being the trace of the gauge potential while $\Box=\partial^\mu\partial_\mu$ is the d'Alambert operator. Imposing, $\delta_dS^h=0$, a straightforward calculation gives the constraints $a_2=-a_1$, $a_3=-2a_1$, and $a_4=2a_1$ over the coefficients. Thus,
\begin{equation}
    S^h_{lin}=\frac{a}{2}\int d^4x\left(-h\Box h+h^a_\mu\Box h^\mu_a-2h^\mu_a\partial_\mu\partial^\nu h^a_\nu+2h\eta^\mu_a\partial_\mu\partial^\nu h^a_\nu\right)\;,\label{acth2}
\end{equation}
where we have redefined $a_1=-a/2$ for convenience. This is the usual Fierz-Pauli massless action \cite{Fierz:1939ix,Gambuti:2021meo} in the EC formalism, \emph{i.e.} the non-interacting theory for spin-2 massless gravitons.

The second exercise is to follow the ideas in \cite{Bertolini:2022ijb,Bertolini:2023juh}, for the fracton theory. In essence, it can be obtained from a more restricted class of the gauge transformations \eqref{diff2} by setting $\zeta^\alpha=\partial^\alpha\zeta$. Let us call it fracton gauge symmetry and define $\delta_f$ by 
\begin{eqnarray}
\delta_fh^a_\mu&=&\eta^a_\nu\partial^\nu\partial_\mu\zeta\;,\nonumber\\
\delta_fh_a^\mu&=&\eta_a^\nu\partial_\nu\partial^\mu\zeta\;.\label{diff3}
\end{eqnarray}
Thence, if we now impose fracton symmetry to action \eqref{acth1}, $\delta_fS^h=0$, one straightforwardly obtains the constraints $a_3=-2a_1$ and $a_2=a_1-a_4$. Hence, $S^h$ reduces to
\begin{equation}
    S^h=\int d^4x\left[b_1\left(-h\Box h+2h\eta^\mu_a\partial_\mu\partial^\nu h^a_\nu-h^\mu_a\partial_\mu\partial^\nu h^a_\nu\right)+b_2\left(-h^a_\mu\Box h^\mu_a+h^\mu_a\partial_\mu\partial^\nu h^a_\nu\right)\right]\;,\label{acth3}
\end{equation}
where we have redefined $a_1\rightarrow -b_1$ and $a_2\rightarrow -b_2$. Performing two more redefinitions, $b_1\rightarrow a/2$ and $b_2\rightarrow(b-a)/2$, we finally get
\begin{equation}
S^h=S^h_{lin}+S^h_{frac}\;,\label{actfrac1}
\end{equation}
with
\begin{equation}
    S^h_{frac}=\frac{b}{2}\int d^4x\left(-h^a_\mu\Box h^\mu_a+h^\mu_a\partial_\mu\partial^\nu h^a_\nu\right)\;,\label{actfrac2}
\end{equation}
being the action of fractons in terms of the EC variables $h^a_\mu$. Action \eqref{actfrac1} is just the EC version of the BM model \cite{Blasi:2022mbl,Bertolini:2022ijb,Bertolini:2023juh} (See Sect.~\ref{BBDM2}). From the gauge construction of the model, the parameters $a$ and $b$ are arbitrary constants (with dimension of mass squared). Nevertheless, $a$ can be related with Newton's constant if action \eqref{actfrac1} refers to some kind of gravity model, by means of $a=1/16\pi G$. In fact, as discussed in \cite{Bertolini:2023juh}, pure linear gravity theory is achieved by setting $b=0$ while pure fractonic action is obtained for $a=0$. For any other non trivial values of these parameters, a mixed model of fractons and gravitons is at our disposal. Remarkably, in the case of $a=0$, there is a regime where the field equations split into ordinary Maxwell equations for scalar fracton charges and $SO(3)$ algebra-valued fractonic electrodynamics (See Sect.~\ref{MAX}).

It is useful to rewrite action \eqref{actfrac1} in terms of the gauge invariant field $H_{\mu\nu}^a$,
\begin{equation}
    S^h=\frac{a}{2}\int d^4x H_\mu H^\mu+\frac{(b-a)}{4}\int d^4x H^a_{\mu\nu}H_a^{\mu\nu}\;,\label{actfrac1a}
\end{equation}
with
\begin{eqnarray}
    H^a_{\mu\nu}&=&\partial_\mu h^a_\nu-\partial_\nu h^a_\mu\;,\nonumber\\
    H_\mu&=&\eta_a^\nu H_{\mu\nu}^a\;.\label{hh2}
\end{eqnarray}
And, to complete the model, at least classically, we couple action \eqref{actfrac1a} with an external source field\footnote{For a covariant coupling with a complex scalar field, see \cite{Sobreiro:2023rsu}.} $\tau^\mu_a$, so the total action of interest reads
\begin{equation}
    S=S^h-\int d^4x\;h^a_\mu\tau^\mu_a\;.\label{acttot1}
\end{equation}
The source $\tau^\mu_a$ is regarded as a covariant current of fractonic charges. The corresponding field equations generated from the action \eqref{acttot1} are easily computed,
\begin{equation}
    \partial_\nu\left[(b-a)H^{\mu\nu}_a+a\left(\eta^\mu_aH^\nu-\eta^\nu_aH^\mu\right)\right]=\tau^\mu_a\;.\label{feq1a}
\end{equation}
Taking the divergence of the field equations \eqref{feq1a}, one directly obtains the fracton charge conservation equation
\begin{equation}
\partial_\mu\tau^\mu_a=0\;.\label{cons1}
\end{equation}
Equation \eqref{feq1a} can be simplified a bit by taking its trace, yielding
\begin{equation}
\partial_\mu H^\mu=\frac{1}{(b+2a)}\tau\;.
\end{equation}
Thus, \eqref{feq1a} reduces to
\begin{equation}
    \partial_\nu\left[(b-a)H^{\mu\nu}_a-a\eta^\nu_aH^\mu\right]=\theta^\mu_a\;,\label{feq1b}
\end{equation}
where
\begin{equation}
    \theta^\mu_a=\tau^\mu_a-\frac{a}{(b+2a)}\eta^\mu_a\tau\;,
\end{equation}
is an effective current whose trace is given by
\begin{equation}
    \theta=\frac{(b-2a)}{(b+2a)}\tau\;.
\end{equation}
Clearly, if $b=2a$, the effective current is traceless. Although equation \eqref{feq1b} is simpler than \eqref{feq1a}, the effective current is no longer a conserved quantity.

Besides the field equations \eqref{feq1a}, the Bianchi identity is also valid
\begin{equation}
\epsilon^{\mu\nu\alpha\beta}\partial_\nu H_{\alpha\beta}^a=0\;.\label{bia1}
\end{equation}
The Bianchi identities complete the set of field equations for $h^a_\mu$. Another way to look, inspired in electrodynamics, is to solve equations \eqref{feq1a} and \eqref{bia1} for the gauge invariant field $H^a_{\mu\nu}$, which is the analogous field of the electromagnetic field $F_{\mu\nu}=\partial_\mu A_\nu-\partial_\nu A_\mu$. 

To end this section, let us define
\begin{equation}
    \mathcal{F}_{\mu\nu}^a=(b-a) H^a_{\mu\nu}+a\left(\eta_\mu^aH_\nu-\eta_\nu^aH_\mu\right)\;,\label{F0}
\end{equation}
which is the tensor field appearing in equation \eqref{feq1a}. Thus,
\begin{equation}
\partial^\nu\mathcal{F}^a_{\mu\nu}=\tau^a_\mu\;,\label{feq1c}
\end{equation}
is clearly the generalization of equation \eqref{frac1}. First, because it is a covariant generalization. Second, due to the $a$ and $b$ parameters characterizing fractonic and gravitational sectors. Third, because the field and source carry a gauge index. In fact, the gauge index will reveal the photonic sector of the model in a certain regime. Nevertheless, there is a difference: the field $H^a_{\mu\nu}$ is antisymmetric in spacetime indices while $E_{ij}$ is symmetric. Indeed, our construction naturally led to an antisymmetric representation of the gauge invariant fractonic fields. In Section \ref{BBDM2}, we will prove that this representation is equivalent to the symmetric representation of the BM model.

\section{Maxwell electrodynamics and SO(3) algebra-valued electrodynamics}\label{MAX}

In this section, we turn off the gravitational channel by setting the parameter $a$ to zero and assume a pure covariant fracton electrodynamical theory with gauge index. The field equations \eqref{feq1a} reduce to
\begin{equation}
    \partial_\nu H_a^{\mu\nu}=\tau^\mu_a\;.\label{feq2}
\end{equation}
where the constant $b$ has been absorbed in $h^a_\mu$ by means of the rescaling $h^a_\mu\rightarrow b^{-1}h^a_\mu$. 

Recalling that $H^a_{\mu\nu}$ is gauge invariant, and thus observable, and that the Bianchi identity \eqref{bia1} is always valid, we have a set of equations very similar to Maxwell equations. Except, of course, for the gauge index. The gauge index, referring to the tangent space, is an internal index analogous to color indices in quantum chromodynamics \cite{Itzykson:1980rh}, a genuine gauge charge index. However, inhere, by construction, the gauge index refers to a gauge symmetry identified with local spacetime isometries characterized by the Lorentz group of transformations connecting all local inertial frames. Therefore, the set of equations \eqref{feq2} and \eqref{bia1} describe an $SO(1,3)$ algebra-valued theory with gauge symmetry \eqref{diff3}: The fields and charges are $SO(1,3)$ algebra-valued objects and they transform under gauge transformations \eqref{diff3}. These gauge transformations can be cast in the form
\begin{eqnarray}
\delta_fh^a_\mu&=&\partial_\mu\zeta^a\;,\nonumber\\
\delta_fh_a^\mu&=&\partial^\mu\zeta_a\;,\label{diff3a}
\end{eqnarray}
with $\zeta^a=\eta^a_\nu\partial^\nu\zeta$, bringing the model more closely related to the standard EC theory of gravity.

At this point, we explore the fact that the gauge indices are tangent space indices: We impose a restrictive condition to the frames $x^a$: We fix the time axis and leave an $SO(3)$ global gauge freedom for tangent indices. Physically, we are imposing that all frames $x^a$ have synchronized clocks and they relate to each other only by spatial rotations, \emph{i.e.} they do not move with respect to each other. 

We consider first the index $a$ projected in time direction, $a=\underline{0}$. Since we have fixed this axis, these components do not transform under the residual $SO(3)$ possible rotations. Therefore, in this regime we are allowed to omit all indices $a=\underline{0}$. Remarkably, in this case, we recover the equations of ordinary electrodynamics. In fact, identifying $h^{\underline{0}}_\mu=A_\mu$ with the electrodynamical potential, we have $H^{\underline{0}}_{\mu\nu}=\partial_\mu A_\nu-\partial_\nu A_\mu=F_{\mu\nu}$ being identified with the electromagnetic field. Moreover, we can identify $\tau^{\underline{0}}_\mu=j_\mu$ with electric charge four-current ($j_0=\rho$ is the scalar fractonic charge density and $j_i$ is the fractonic vector current density). Consequently, it is trivial to show that \eqref{feq2} and \eqref{bia1} are exactly reduced to Maxwell equations. Furthermore, electrodynamics is a realization of a $U(1)$ gauge theory. Indeed, gauge transformations \eqref{diff3} (or \eqref{diff3a}) reduce to $U(1)$ gauge transformations,
\begin{eqnarray}
\delta_eA_\mu&=&\partial_\mu\xi\;,\nonumber\\
\delta_eA^\mu&=&\partial^\mu\xi\;,\label{diff4}
\end{eqnarray}
with $\xi=\eta^{\underline{0}}_\nu\partial^\nu\zeta$ and $\delta_f\rightarrow\delta_e$. Hence, in this regime, fractonic charges manifest themselves exactly like electric charges (including their conservation law, $\partial_\mu j^\mu=0$, coming from \eqref{cons1}) and the same occurs to the fractonic gauge fields in this sector. Essentially, this regime has no apparent difference from ordinary electrodynamics. 

For $a=\underline{i}$, we define $H^{0i}_{\underline{i}}=E^i_{\underline{i}}$, $H^{ij}_{\underline{i}}=\epsilon^{ij}_{\phantom{ij}k}B^k_{\underline{i}}$, $\tau^0_{\underline{i}}=\rho_{\underline{i}}$ and $\tau^i_{\underline{i}}=j^i_{\underline{i}}$, and equations $\eqref{feq2}$ become
\begin{eqnarray}
\vec{\nabla}\cdot\vec{E}_{\underline{i}}&=&\rho_{\underline{i}}\;.\nonumber\\
    \vec{\nabla}\times\vec{B}_{\underline{i}}-\frac{\partial\vec{E}_{\underline{i}}}{\partial t}&=&\vec{j}_{\underline{i}}\;.\label{feq4}
\end{eqnarray}
Thus, we attained an $SO(3)$ algebra-valued electrodynamics. Electric and magnetic fields are vector fields, but they carry a gauge index associated with the global $SO(3)$ rotations in tangent space. The gauge transformations \eqref{diff3}, on the other hand, become
\begin{eqnarray}
\delta_ph^{\underline{i}}_\mu&=&\partial_\mu\xi^{\underline{i}}\;,\nonumber\\
\delta_ph_{\underline{i}}^\mu&=&\partial^\mu\xi_{\underline{i}}\;,\label{diff5}
\end{eqnarray}
with $\xi^{\underline{i}}=\eta^{\underline{i}}_\nu\partial^\nu\zeta$ and $\delta_f\rightarrow\delta_p$. Transformations \eqref{diff5} describe a local gauge symmetry, classifying this sector of the model as a kind of triple electrodynamics whose conserved charges are $SO(3)$ algebra-valued objects.

Let us go back to the full covariant equation \eqref{feq2} and the Bianchi identity \eqref{bia1}. We have a generalized covariant fracton model where the four-current $\tau^\mu_a$ describes a generalization of fractonic vector charges. And, for a specific relativistic regime, the model splits in a sector indistinguishable from ordinary electrodynamics and a triple electrodynamics whose charges and fields are $SO(3)$ algebra-valued objects. We conclude therefore that the full model \eqref{acttot1} carries, not only fractonic and gravitonic modes, but also photonic modes. These new modes are indeed present in the general theory \eqref{acttot1}, but, in this case, they can be distinguished from ordinary electromagnetic physics by performing a Lorentz transformation in tangent space.

\section{Fracton electrodynamics in fully spacetime representation}\label{BBDM2}

We have constructed the model \eqref{acttot1} based on the same gauge principles of the BM model \cite{Blasi:2022mbl,Bertolini:2022ijb,Bertolini:2023juh}, except for the basic field which is not a symmetric tensor but a vector field with an internal charge index. We have attained a model of gauge invariant antisymmetric fractonic fields. Therefore, we still have to show how our model relates to the BM model, which is constructed in terms of symmetric tensor fields.

To connect our model with the BM model, we must perform a change of variables from $h^a_\mu$ to a fully spacetime representation $h_{\mu\nu}$. It turns out that the connection between EC variables $h^a_\mu$ and second order variables $h_{\mu\nu}$ is given simply by
\begin{equation}
    h_{\mu\nu}=\frac{1}{2}\left(\eta_{a\mu}h^a_\nu+\eta_{a\nu}h^a_\mu\right)\;.\label{hh1}
\end{equation}
It is then straightforward to show that
\begin{equation}
    F_{\mu\nu\alpha}=\eta_{a\mu}H^a_{\nu\alpha}+\eta_{a\nu}H^a_{\mu\alpha}=\partial_\mu h_{\alpha\nu}+\partial_\nu h_{\mu\alpha}-2\partial_\alpha h_{\mu\nu}\;,\label{F1}
\end{equation}
which is the fracton gauge invariant field strength defined in \cite{Bertolini:2022ijb}. In fact, applying $\partial^\alpha$ in \eqref{F1} and using the field equations \eqref{feq2} one readily finds equations \eqref{frac2},
where the fracton current in spacetime representation reads
\begin{equation}
    \tau_{\mu\nu}=\eta_{a\mu}\tau^a_\nu+\eta_{a\nu}\tau^a_\mu\;.\label{source1}
\end{equation}
The field equation \eqref{frac2} is the pure fracton equation (with sources) of the BM model. The full equation in the BM model actually reads
\begin{equation}
    (b-a)\partial^\alpha F_{\mu\nu\alpha}+a\left[\eta_{\mu\nu}\partial_\alpha F_\beta^{\phantom{\beta}\beta\alpha}-\frac{1}{2}\left(\partial_\mu F_\nu+\partial_\nu F_\mu\right)\right]=\tau_{\mu\nu}\;,\label{feq6}
\end{equation}
with $F_\mu=F^\nu_{\phantom{\nu}\nu\mu}$. It is a straightforward exercise to show that the substitution of \eqref{F1} and \eqref{source1} in the field equations \eqref{feq6} yields equations \eqref{feq1a}. An easier task is to show that action \eqref{acttot1}, in terms of the field strength \eqref{F1}, reads
\begin{equation}
    S^h(F)=\frac{(b-a)}{12}\int d^4x F_{\mu\nu\alpha}F^{\mu\nu\alpha}+\frac{a}{8}\int d^4x F^\nu_{\phantom{\nu}\nu\mu}F^\alpha_{\phantom{\alpha}\alpha\mu}\;,
\end{equation}
which is, up to normalization factors, the BM action proposed in \cite{Bertolini:2022ijb}. This proves that the model described by action \eqref{acttot1} is totally equivalent to the BM model \cite{Bertolini:2022ijb}.

\section{Conclusions}\label{conc}

In this paper we constructed a covariant model of fractons based on purely gauge principles and using EC variables in complete analogy to the BM gauge model of fractons \cite{Blasi:2022mbl,Bertolini:2022ijb,Bertolini:2023juh}. The general action, quadratic on the gauge field, is composed of a pure fractonic sector and another sector identified as the Fierz-Pauli action for massless gravitons. We proved, by changing the EC variables to second-order variables, that our model is totally equivalent to the BM model, as expected. In terms of gauge invariant fields, we have constructed a fractonic model in a representation of antisymmetric fields, in contrast to the BM model of symmetric fields. Nevertheless, we proved the equivalence between both models.

Additionally, we showed that there is a specific relativistic regime where Maxwell electrodynamics emerges. In this regime, observers with synchronized clocks, not moving with respect to each other, cannot distinguish this sector of fracton electrodynamics from ordinary Maxwell electromagnetism. This effect also works for non-relativistic observers, \emph{i.e.}, a relativistic regime where absolute time can be defined for local observers (\emph{e.g.} Galilean frames \cite{Guerrieri:2020vhp}). Moreover, the $SO(3)$ algebra-valued vector charges and fields are also present in this regime, a kind of triple electrodynamics.

It turns out that the relativistic regime favoring fractons to mimic electrodynamics is just the typical situation occurring in most solid state laboratories. Therefore, such an effect could be probed somehow by observing, for instance, a kind of phase in a spin liquid looking like an electrically charged spin liquid. Such a system would produce fake electromagnetic fields and interact with them. However, it is expected that the system would not respond to ordinary electromagnetic fields normally because the charges are actually fracton charges and not electrical ones. Therefore, it seems that the effect proposed here is actually measurable under laboratory conditions.

The results obtained can be improved by exploring further the model in EC representation. For instance, one can couple scalar fields instead of an external current, providing dynamics to the fracton charges. Another possibility is to couple fermions, since the EC formalism is a suitable scenario to couple gravity with spinors \cite{DeSabbata:1986sv,Hehl:1994ue}.

\section*{Acknowledgements}

This study was financed in part by The Coordena\c c\~ao de Aperfei\c coamento de Pessoal de N\'ivel Superior - Brasil (CAPES) - Finance Code 001.

\bibliography{BIB}
\bibliographystyle{utphys2}

\end{document}